\documentclass[aps,twocolumn,superscriptaddress]{revtex4-2}
\usepackage[T1]{fontenc}
\usepackage{graphicx}
\usepackage{amssymb,amsmath}
\usepackage[colorlinks=true, citecolor={blue!80!black}, urlcolor={blue!80!black}, linkcolor = {blue!80!black}]{hyperref}
\usepackage{braket}
\usepackage{placeins}
\usepackage{tabularx}
\usepackage[separate-uncertainty=true]{siunitx}
\usepackage{physics}
\usepackage{makecell}
\usepackage{appendix}
\usepackage{soul}
\AtBeginDocument{\RenewCommandCopy\qty\SI}

\usepackage{pifont}
\usepackage{mathtools}
\makeatletter
\newcommand*{\addFileDependency}[1]{% argument=file name and extension
\typeout{(#1)}% latexmk will find this if $recorder=0
\@addtofilelist{#1}
\IfFileExists{#1}{}{\typeout{No file #1.}}
}\makeatother

\newcommand{\Sb}{^{123}\rm Sb}
\newcommand{\Ge}{^{73}\rm Ge}

\usepackage{xr}

\usepackage{xcolor}
\usepackage{soul}

\begin{document}
%TC:ignore

\title{Maximizing the nondemolition nature of a quantum measurement via an adaptive readout protocol}

% \title{Minimizing the impact of non-ideal QND measurements with an adaptive readout protocol}

% \title{Minimizing measurement-induced errors with an adaptive readout protocol}

\author{Arjen Vaartjes}
\affiliation{School of Electrical Engineering and Telecommunications, UNSW Sydney, Sydney, NSW 2052, Australia}
\author{Rocky Yue Su}
\affiliation{School of Electrical Engineering and Telecommunications, UNSW Sydney, Sydney, NSW 2052, Australia}
\author{Laura A. O'Neill}
\affiliation{School of Electrical Engineering and Telecommunications, UNSW Sydney, Sydney, NSW 2052, Australia}
\author{Paul Steinacker}
\affiliation{School of Electrical Engineering and Telecommunications, UNSW Sydney, Sydney, NSW 2052, Australia}
\affiliation{Diraq Pty. Ltd., Sydney, NSW, Australia}
\author{Gauri Goenka}
\affiliation{School of Electrical Engineering and Telecommunications, UNSW Sydney, Sydney, NSW 2052, Australia}
\author{Mark R. van Blankenstein}
\affiliation{School of Electrical Engineering and Telecommunications, UNSW Sydney, Sydney, NSW 2052, Australia}
\author{Xi Yu}
\affiliation{School of Electrical Engineering and Telecommunications, UNSW Sydney, Sydney, NSW 2052, Australia}
\author{Benjamin Wilhelm}
\affiliation{School of Electrical Engineering and Telecommunications, UNSW Sydney, Sydney, NSW 2052, Australia}
\author{Alexander M. Jakob}
\affiliation{School of Physics, University of Melbourne, Melbourne, VIC 3010, Australia}
\author{Fay E. Hudson}
\affiliation{School of Electrical Engineering and Telecommunications, UNSW Sydney, Sydney, NSW 2052, Australia}
\affiliation{Diraq Pty. Ltd., Sydney, NSW, Australia}
\author{Kohei M. Itoh}
\affiliation{School of Fundamental Science and Technology, Keio University, Kohoku-ku, Yokohama, Japan}
\author{Chih Hwan Yang}
\affiliation{School of Electrical Engineering and Telecommunications, UNSW Sydney, Sydney, NSW 2052, Australia}
\affiliation{Diraq Pty. Ltd., Sydney, NSW, Australia}
\author{Andrew S. Dzurak}
\affiliation{School of Electrical Engineering and Telecommunications, UNSW Sydney, Sydney, NSW 2052, Australia}
\affiliation{Diraq Pty. Ltd., Sydney, NSW, Australia}
\author{David N. Jamieson}
\affiliation{School of Physics, University of Melbourne, Melbourne, VIC 3010, Australia}
\author{Martin Nurizzo}
\affiliation{School of Electrical Engineering and Telecommunications, UNSW Sydney, Sydney, NSW 2052, Australia}
\author{Danielle Holmes}
\affiliation{School of Electrical Engineering and Telecommunications, UNSW Sydney, Sydney, NSW 2052, Australia}
\author{Arne Laucht}
\affiliation{School of Electrical Engineering and Telecommunications, UNSW Sydney, Sydney, NSW 2052, Australia}
\affiliation{Diraq Pty. Ltd., Sydney, NSW, Australia}
\author{Andrea Morello}\thanks{Lead contact: \url{a.morello@unsw.edu.au}}
\affiliation{School of Electrical Engineering and Telecommunications, UNSW Sydney, Sydney, NSW 2052, Australia}

\begin{abstract}
 
Quantum error correction (QEC) requires non-invasive measurements for fault tolerant quantum computing. Deviations from ideal quantum non-demolition (QND) measurements can disturb the encoded information. To address this challenge, we develop a readout protocol for a $D-$dimensional system that, after a single positive outcome, switches to probing only the $D{-}1$ remaining subspace. This adaptive switching strategy minimizes measurement-induced errors by relying on negative-result measurement results that do not perturb the Hamiltonian. We apply the protocol on an 8-dimensional $\Sb$ nuclear qudit in silicon, and achieve an increase in the readout fidelity from \SI{98.93\pm0.07}{\percent} to \SI{99.61\pm0.04}{\percent}, while reducing threefold the overall  readout time. To highlight the broader relevance of measurement-induced errors, we study a 10-dimensional $\Ge$ nuclear spin read out through Pauli spin blockade, revealing nuclear spin flips arising from hyperfine and quadrupole interactions. These results unveil the effect of non-ideal QND readout across diverse platforms, and introduce an efficient readout protocol that can be implemented with minimal FPGA logic on existing hardware. 

\end{abstract}

\maketitle

%TC:endignore
\section{Introduction}

Quantum error correction (QEC) \cite{devitt2013quantum} relies on high-fidelity measurements to extract syndromes without introducing additional non-correctable errors. 
Quantum non-demolition (QND) measurements \cite{braginsky1996quantum} are essential in this context because they allow repeated readout to improve confidence in the measured observable. Non-demolition does not imply the absence of measurement backaction - indeed, any projective measurement inevitably collapses the wavefunction onto an eigenstate of the measurement operator. However, in the ideal QND case, this backaction does not accumulate when the measurement is repeated, and the system remains in the same eigenstate it was projected onto after the first shot. This principle underpins stabilizer measurements in QEC and allows for multiple cycles of error correction \cite{Riste2015DetectingMeasurements, Kelly2015StateCircuit}.

A measurement is perfectly QND if the eigenstates of the measurement operator are identical to the eigenstates of the system Hamiltonian. In ancilla-mediated QND readout, this condition is met when the system Hamiltonian $H_S$ commutes with the Hamiltonian describing the coupling to the ancilla $H_C$:~\cite{unruh1979quantum,braginsky1996quantum,Joecker2024}
\begin{equation}
    [H_C, H_S] = 0.
    \label{eq:commutation}
\end{equation}

If condition~(\ref{eq:commutation}) is not met, however, the coupling to the ancilla causes a change in eigenbasis of the quantum system. The basis mismatch between the bare system and the coupled system is what we refer to as non-ideal QND behavior and causes transitions between eigenstates upon repeated measurement of the quantum state.

We focus here on the case where QND readout of a quantum system is achieved by coupling it to an ancilla, which is subsequently and separately measured. This scenario has been employed in several platforms such as spin qubits in gate-defined quantum dots \cite{Xue2020, Yoneda2020}, NV-center spins in diamond \cite{Neumann2010, Robledo2013}, trapped ions \cite{Negnevitsky2018} and nuclear spins \cite{vincent2012electronic,Pla2013, Asaad2020, Fernandez2024, Reiner2024, Stolte2025}. In the case of nuclear spin qubits (with nuclear spin $I=1/2$, Hilbert space dimension $D=2$) \cite{Pla2013, Reiner2024} or qudits ($I\geq 1$, $D>2$) \cite{vincent2012electronic, Asaad2020, Fernandez2024, Stolte2025}, the ancilla is formed by a hyperfine-coupled electron. The nuclear spin measurement proceeds by first making the electron spin state depend on the nuclear one - usually via a selective spin-inversion pulse - and then reading out the electron ancilla. For the case where readout is performed electrically, the electron is spin-dependently \cite{Elzerman2004Single-shotDot,xiao2004electrical,Morello2010} allowed to tunnel to the island of a single-electron transistor \cite{morello2009architecture, Morello2010, Pla2013}, or to a nearby quantum dot (QD) \cite{Geng2025}, or through a spin-polarized scanning tunneling microscope tip \cite{Stolte2025}. Upon tunneling, the coupling Hamiltonian formed by the hyperfine interaction suddenly decreases to zero, which perturbs the eigenstates of the nuclear spin. As a result, the nuclear state will be projected onto a slightly modified eigenbasis which probabilistically causes the nuclear spin to flip. This effect is known as ionization shock \cite{Pla2013, Hile2018AddressableSilicon, Madzik2022PrecisionSilicon}. It is usually possible to quantify the deviation from ideal QND measurement by counting, on average, how many electron ionization events are necessary to induce an accidental flip of the nuclear state \cite{Pla2013,Madzik2022PrecisionSilicon}. 

In this work, we demonstrate an adaptive readout protocol that improves the QND readout fidelity of a $D=8$ nuclear qudit by minimizing the number of electron ionization events. Instead of performing repeated QND readout cycles over the full $D$-dimensional state space of a quantum system, after a single successful readout event the protocol adaptively switches to measuring the remaining $D-1$ states, which we refer to as the `dark state subspace'. The electron spin readout is a type of `negative-result measurement'  \cite{elitzur1993quantum}, where one of the outcomes is associated with the absence of tunneling \cite{Muhonen2018}. Therefore, probing the dark state subspace allows the extraction of information without paying the penalty of measurement-induced spin flips.
In the ideal limit of perfect ancilla readout, a negative measurement outcome within the dark state subspace is truly QND, because the Hamiltonian remains unchanged throughout the process. Using the adaptive readout protocol, we report an increase in the average readout fidelity from \SI{98.93\pm0.07}{\percent}  to \SI{99.61\pm0.04}{\percent}, compared to regular repeated QND readout, while achieving a factor 3 speed-up.

\begin{figure*}
    \centering
    \includegraphics[width=\textwidth]{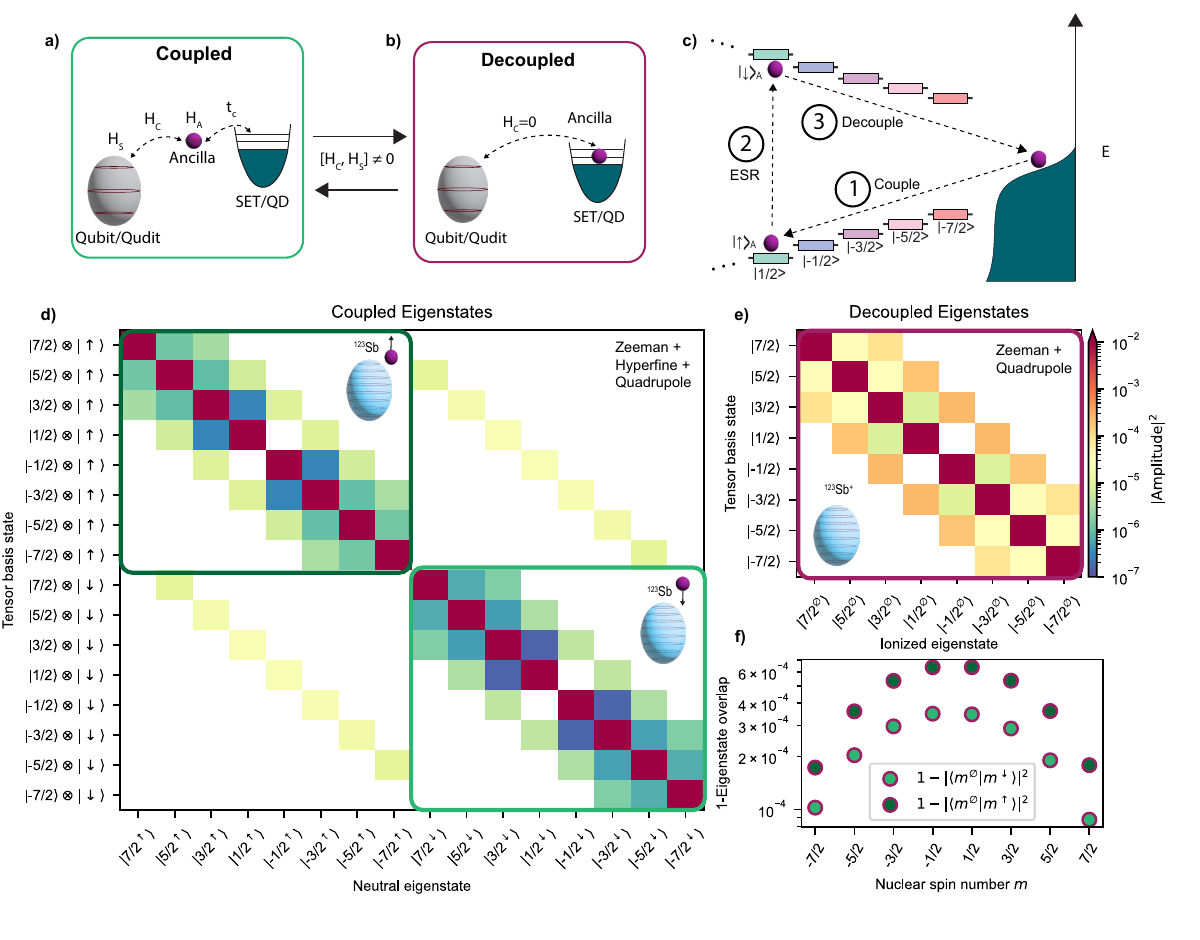}
    \caption{\textbf{Deviation from ideal QND measurements due to an eigenbasis mismatch of coupled and decoupled systems}. \textbf{a)} In the coupled system, the qubit or qudit is coupled to an ancilla, which is tunnel coupled to a reservoir or quantum dot (QD).
    \textbf{b)} If the ancilla tunnels onto the reservoir/QD, the coupling Hamiltonian $H_c$ reduces to 0. This process induces spin flips if $[H_S, H_C]\neq0$. 
    \textbf{c)} One QND cycle consists of coupling, conditional Electron Spin Resonance (ESR), and decoupling the ancilla. 
    \textbf{d)} The eigenstates of the coupled Hamiltonian do not fully overlap with the tensor basis due to the quadrupole and hyperfine interaction. The quadrupole shows up as $\Delta m=\pm1, 2$ off-diagonal elements in this matrix, whereas the hyperfine enables coupling between nuclear spin $\Delta m=\pm1$ and electron spin $\Delta m_e=\mp 1$ components, visible as off-diagonal elements in the top left and bottom right quadrants. 
    \textbf{e)} In the decoupled case, the hyperfine interaction reduces to zero ($H_C=0$) and the influence of the quadruople becomes more pronounced.
    \textbf{f)} The overlap of decoupled and coupled eigenstates - $|{m^\emptyset}\rangle$ and $|m^{\updownarrow}\rangle$ respectively - when the ancilla spin state is $\ket{\downarrow}$ (light green) or $\ket{\uparrow}$ (dark green).}
    \label{fig:1}
\end{figure*}

\section{Methods}

We demonstrate the adaptive readout protocol on a quadrupolar spin-7/2 $\Sb$ nucleus implanted in isotopically enriched silicon, forming an 8-dimensional qudit \cite{Asaad2020, Fernandez2024}. A hyperfine-coupled electron acts as an ancilla for QND readout and is tunnel-coupled to a nearby charge reservoir (see Fig.\ref{fig:1}a). The measurement-induced dynamics of the system is governed by the Hamiltonians of the qudit, the ancilla, and their coupling, which we describe below.

\subsection{System Hamiltonian and Measurement-induced Dynamics}

In the presence of a magnetic field $B_0$, the $\Sb$ donor is described by the Hamiltonian:

\begin{equation}
H = H_S + H_A + H_C.
\label{eq:full_hamiltonian}
\end{equation}

Here, the Hamiltonian of the qudit system and ancilla are given by: 
\begin{equation}
    H_S=-\gamma_n B_0I_z + \sum_{\alpha, \beta}Q^+_{\alpha\beta}I_\alpha I_\beta, 
    \label{eq:bare_hamiltonian}
\end{equation}
\begin{equation}
    H_A=\gamma_e B_0 S_z,
    \label{eq:ancilla_hamiltonian}
\end{equation}
and the hyperfine coupling is defined as:
\begin{equation}
    H_C=A \vec{I}\cdot \vec{S}
    \label{eq:coupling_hamiltonian}
\end{equation}
The definitions of the Hamiltonian parameters are summarized in Tab.\ref{tab:hamiltonian parameters}. 
\begin{table}[h]
\centering
\begin{tabular}{ll}
\hline
\textbf{Symbol} & \textbf{Description} \\
\hline
$B_0=\SI{1.395}{\tesla}$ & External magnetic field \\
$\gamma_n=\SI{5.55}{\mega\hertz/\tesla}$ & Nuclear gyromagnetic ratio \\
$\gamma_e=\SI{28.02}{\giga\hertz/\tesla}$ & Electron gyromagnetic ratio \\
$A=\SI{97.5}{\mega\hertz}$ & Hyperfine interaction \\
$I = [I_x, I_y, I_z]$ & Nuclear spin-7/2 operators ($8 \times 8$ matrices) \\
$S = [S_x, S_y, S_z]$ & 
Electron spin-1/2 operators ($2 \times 2$ matrices) \\
$\alpha, \beta$ & Cartesian indices: $x$, $y$, $z$ \\
$Q^+_{\alpha\beta}$ & \makecell[l]{Quadrupole tensor element in the decoupled \\ (ionized) case} \\
\hline
\end{tabular}
\caption{Physical parameters and operators in the system}
\label{tab:hamiltonian parameters}
\end{table}

Equation (\ref{eq:coupling_hamiltonian}) clarifies the origin of measurement-induced spin flips in the $\Sb$ system. The hyperfine coupling Hamiltonian does not commute with either the Zeeman term ($I_z$) or with the quadrupolar terms ($I_\alpha I_\beta$) in $H_S$. As a result, we experience a change in eigenstates upon coupling and decoupling the ancilla during QND readout. Due to the presence of both linear and quadratic spin operators, both first-order ($\Delta m=\pm1$) and second-order ($\Delta m=\pm2$) transitions will be present during QND readout, where $\Delta m$ indicates the change in nuclear spin number. 

% To verify physical consistency, we simulate the electric field gradients (EFG) that give rise to the quadrupolar interaction in COMSOL using the device layout (see Supplementary~X). The fitted tensor is consistent with the simulated EFG field and also enables precise triangulation of the donor position \cite{Oneill2025}.

\subsection{QND readout cycle}

Figure.~\ref{fig:1}d outlines the steps of a QND readout cycle for the $\Sb$ qudit (see also Supplemental Methods 1 of Ref.\cite{Vaartjes2025}). The ancilla electron is denoted by the spin states $\ket{\uparrow}$ or $\ket{\downarrow}$ when coupled to the nucleus, and $\ket{\emptyset}$ when decoupled. The nuclear spin states are described in the basis of the eigenstates of $I_z$, $\{\ket{m}\}$, with $m=7/2, 5/2, \ldots, -7/2$.

In Step 1, we induce the transition $\ket{\emptyset}\rightarrow\ket{\downarrow}$. An electron is loaded onto the nucleus from a nearby Single Electron Transistor (SET) by applying a pulse in local gate voltages \cite{Morello2010}. We ensure that the loaded electron is truly in the $\ket{\downarrow}$ state with high fidelity by using a \textit{Bayesian Maxwell's demon} approach, introduced in Ref. \cite{Johnson2022}. 

In Step 2, we adiabatically flip the spin of the electron $\ket{\downarrow} \rightarrow \ket{\uparrow}$, conditional on the nuclear spin state using an adiabatic Electron Spin Resonance (aESR) pulse \cite{Laucht2014}, consisting of a frequency chirp that envelops the desired resonance frequency. This can be either a single selective pulse or a multi-frequency pulse targeting a specific nuclear subspace~\cite{Yu2025}.

In Step 3, if the aESR pulse flipped the spin to $\ket{\uparrow}$, the electron is allowed to decouple from the nucleus and tunnel back onto the SET, i.e. $\ket{\uparrow} \rightarrow \ket{\emptyset}$. The positively-charged donor left behind after the electron tunnel event will switch the bias point of the SET from Coulomb blockade (SET current $I_{\rm SET}\approx 0$) to the top of a conductance peak ($I_{\rm SET}\approx 1$~nA). The current remains high until another electron, in the $\ket{\downarrow}$ state, tunnels back onto the SET. If this current blip exceeds a threshold, the electron spin state is labeled as $\ket{\uparrow}$ \cite{Morello2010}. 

To determine the nuclear spin state, we repeatedly cycle through the 8 ESR resonance frequencies (Step 2). From the resulting set of tunnel events, we assign a $\ket{\uparrow}$-probability to each nuclear state, and assign the nuclear state as the one with the highest $\ket{\uparrow}$ probability.
% During one readout \textit{shot}, i.e. one complete cycle through all 8 aESR resonance frequencies, we expect a projective measurement in a single eigenstate corresponding to two tunneling events.

\subsection{Eigenstate overlap}
\label{sec:eigenstate_overlap}

We model nuclear spin flip probabilities by calculating overlaps between the eigenstates of the coupled (neutral donor) and decoupled (ionized donor)  Hamiltonians. We label the eigenstates of the coupled system $\ket{m^{\updownarrow}}$, where $m$ is the dominant nuclear spin projection in each specific eigenstate. In general $\ket{m^{\updownarrow}} \neq \ket{m}\otimes \ket{\updownarrow}$, because the terms contained in the hyperfine and quadrupole Hamiltonians do not commute with $I_z$. Similarly, the eigenstates of the decoupled system, $\ket{m^{\emptyset}}$ do not coincide with the $\ket{m}$ basis states because of the quadrupole interaction. In Fig.~\ref{fig:1}d we then plot the overlap of the coupled eigenstates with the tensor product states $\ket{m}\otimes \ket{\updownarrow}$, and in Fig.~\ref{fig:1}e the overlap of the decoupled eigenstates with the $\ket{m}$ basis states \cite{Monir2024, Joecker2024}.

%Specifically, we calculate the projections of the coupled eigenstates $|\psi_i^\uparrow\rangle$ and $|\psi_i^\downarrow\rangle$ onto the decoupled basis states $|\phi_j^\emptyset\rangle$ as shown in Fig.~\ref{fig:1}c-e \cite{Monir2024, Joecker2024}.

In the coupled case, both hyperfine and quadrupole interactions contribute to the eigenstate composition. The hyperfine term introduces `flip-flop' mixing of electron and nuclear spin components \cite{Savytskyy2023}; i.e., the $|5/2^\uparrow\rangle$ eigenstate includes a small admixture of the $|7/2\rangle \otimes \ket{\downarrow}$ basis state (top right and bottom left quadrant of Fig.~\ref{fig:1}d). The quadrupole interaction adds mixing between nuclear spin states, visible as both first and second off-diagonal elements in Fig.~\ref{fig:1}c, since the quadrupolar interaction contains quadratic terms in the spin operators (e.g. $I_x^2, I_y^2$). Overall, the presence of the vastly dominant electron Zeeman term ($\sim$\SI{40}{\giga \hertz} here) and the associated $I_z S_z$ hyperfine coupling ($\sim$\SI{100}{\mega \hertz}) ensure that the coupled eigenstates remain quite close to the tensor product basis, with cross-couplings well below $10^{-4}$. 

The eigenstates of the decoupled system (Fig.~\ref{fig:1}e) show stronger nuclear mixing, since the quadrupole interaction ($\sim$\SI{20}{\kilo\hertz}) is no longer overwhelmed by the $I_z S_z$ term in the hyperfine coupling, and only competes with the nuclear Zeeman energy ($\sim$\SI{8}{\mega\hertz}). As a consequence, the overlap between coupled and decoupled eigenstates is imperfect.

Figure~\ref{fig:1}f shows that the overlap with the ionized eigenstates is larger for the $|m^\downarrow\rangle$ states, where the quadrupole contribution to the nuclear quantization axis is oriented in the same direction as in the ionized system. In contrast, for the $\ket{m^\uparrow}$ states, the dominant hyperfine component changes sign, effectively flipping the nuclear quantization axis and causing the quadrupole tilt to be in the opposite direction compared to the ionized system. Thus, the decoupling process $\ket{\uparrow}\rightarrow\ket{\emptyset}$ produces a stronger rotation of the nuclear eigenbasis than the coupling process $\ket{\emptyset}\rightarrow \ket{\downarrow}$ (see Fig.~S3).

\section{Results}
\subsection{Jump trace statistics}

\begin{figure}[h]
    \centering
    \includegraphics[width=\columnwidth]{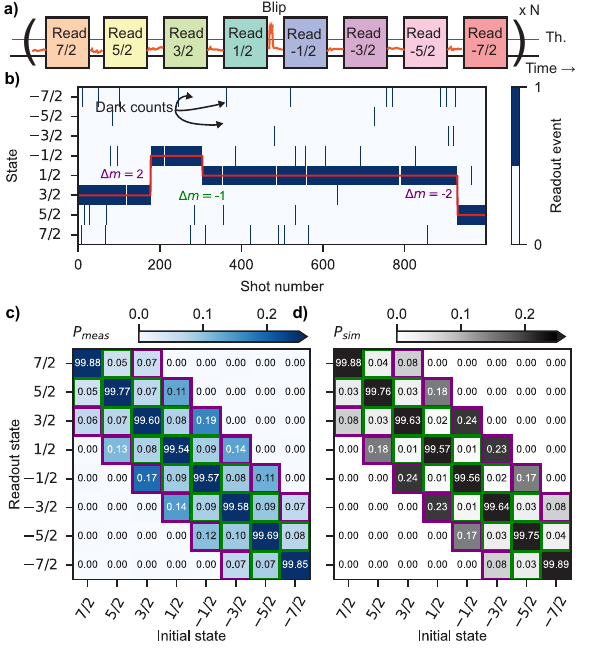}
    \caption{\textbf{Characterization of measurement-induced nuclear spin flips in $\mathbf{\Sb}$} \textbf{a)} Pulse sequence to probe measurement-induced dynamics. We sequentially perform QND readout of the nuclear states $\ket{-7/2}$ through $\ket{+7/2}$, repeated $N$ times. The nuclear state is detected by a short blip of current, caused by the ancillary electron tunneling to a nearby SET. If the blip exceeds the threshold, we count it as a readout event. \textbf{b)} Raw readout results. We determine the most likely state (red line) by window filtering the raw data, to exclude dark counts. We observe the nucleus jumping by either $\Delta m=\pm2$ (purple) or $\Delta m =\pm1$ (green). \textbf{c)} Extracted transition probability matrix. The first and second off-diagonal entries (green and purple) correspond to $\Delta m=1$ and $\Delta m =2$ jumps respectively. \textbf{d)} Simulated transition probability matrix based on eigenstate overlap of the coupled and decoupled nucleus. See Supplemental Materials Section I for model details. }
    \label{fig:2}
\end{figure}

\begin{figure*}[t!]
    \centering
    \includegraphics[width=\textwidth]{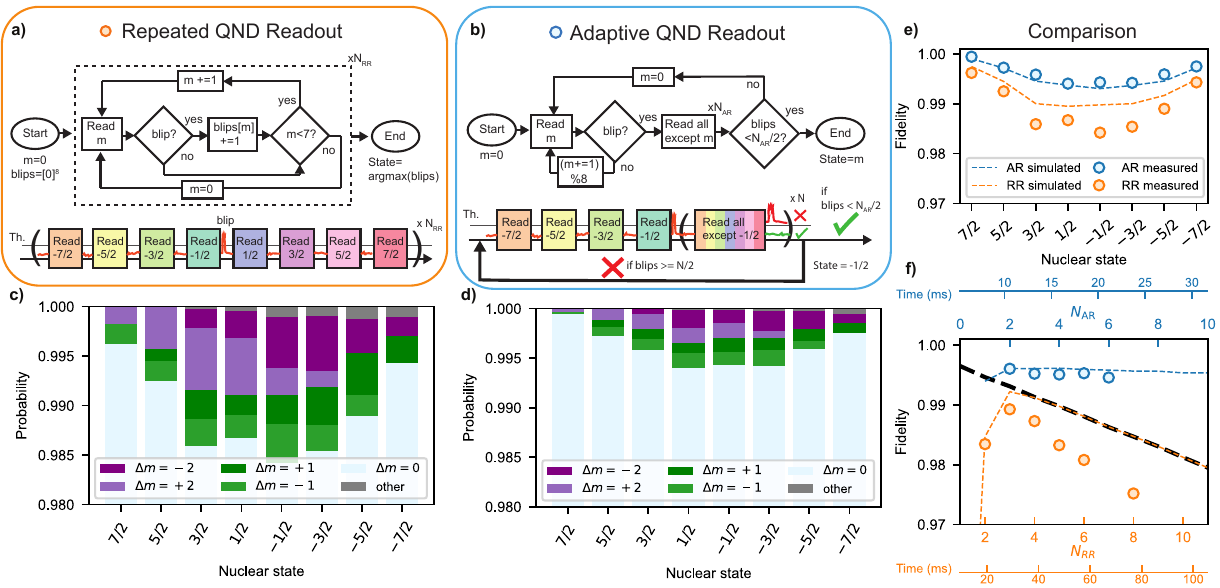}
    \caption{\textbf{Readout improvement using an adaptive QND protocol. a)} Decision tree of conventional repeated QND readout (RR) and corresponding pulse sequence. \textbf{b)} Decision tree of the adaptive QND readout (AR) protocol and an example of a corresponding pulse sequence. As soon as a readout event is detected, the protocol switches to measuring the dark state space $N$ times. \textbf{c)} Nuclear flip probabilities during RR in the best-case scenario. \textbf{d)} Nuclear flip probabilities of AR in the best-case scenario. The adaptive protocol reduces $\Delta m=1$ (green) and $\Delta m=2$ (purple) transition probabilities. \textbf{e)} Comparison of AR (blue) and RR (orange), simulated (lines) and data (dots). The AR data approaches the ionization shock limit for one shot. \textbf{f)} Average readout fidelity comparison vs. number of readout shots and vs. time for RR (orange, bottom axis) and AR (blue, top axis). The black dashed line indicates the simulated RR fidelity if the readout is solely limited by ionization shock (i.e. in the limit of perfect ancilla readout).}
    \label{fig:3}
\end{figure*}

To quantify the measurement-induced dynamics on the $\Sb$ nucleus, we perform a simple repeated readout sequence over all 8 nuclear states, as shown in Fig.~\ref{fig:2}a. 
Figure~\ref{fig:2}b shows an example of raw measurement data, displaying an example of nuclear spin jumps of both $\Delta m=\pm1$ and $\Delta m=\pm2$. Due to ancilla readout imperfections, we observe both dark counts (false positives) and false negatives, which we filter out by applying a kernel-5 majority vote filter.

By analyzing the nuclear jump traces, we extract a transition matrix shown in Fig.~\ref{fig:2}c, which quantifies the nuclear spin flip probabilities during readout. We observe two important features. Firstly, $\Delta m = \pm 2$ transitions dominate over $\Delta m = \pm1$. This means that terms quadratic in the spin operators, i.e. the quadrupole interaction terms, have a dominant effect. Secondly, the flip probability is strongly state dependent, with states at the edges of the Hilbert space ($m = \pm 7/2$) being more resilient against spin flips than states in the middle ($m = \pm 1/2$).

We can qualitatively reproduce the state dependence and dominance of $\Delta m=\pm2$ quadrupolar jumps (Fig.~\ref{fig:2}d) by filling in the Hamiltonian parameters found in equations~(\ref{eq:full_hamiltonian})-(\ref{eq:coupling_hamiltonian}), calculating the eigenstates and computing the transition matrix in equation~(S9).

\subsection{Adaptive readout protocol}

In an ideal QND measurement, increasing the readout repetitions would lead to better readout fidelity. In the presence of deviations from QND ideality, however, each measurement shot poses the risk of altering the measured state. The two phenomena lead to a tradeoff whereby the readout fidelity will reach a maximum as a function of measurement shots, before deteriorating due to QND non-ideality.

%Naively speaking, the more readout repetitions one performs, the better the readout fidelity. However, the measured dynamics illustrate that each tunneling readout event introduces nuclear spin flip probabilities, which decrease the maximal attainable readout fidelity. 

To mitigate this undesirable tradeoff, we implement an adaptive QND readout protocol that minimizes the impact of QND non-ideality. Contrary to the conventional repeated QND readout approach (Fig.~\ref{fig:3}a), the adaptive protocol increases the readout fidelity without the added cost of increasing the amount of Hamiltonian-perturbing tunneling events. 

The protocol is executed in two subroutines (Fig.~\ref{fig:3}b). In the first subroutine, we read each nuclear state sequentially until one tunneling event (blip) is detected. The nuclear state associated with this blip is our initial guess. 
In the second subroutine, instead of repeating QND cycles over all nuclear states - risking multiple tunneling events - we perform a collective measurement of the entire nuclear subspace \emph{excluding} the initial guess. We call this the \textit{dark state subspace}. The collective measurement is implemented by applying 7 ESR pulses while the electron stays coupled - one for each state in the dark state subspace - and only then pulsing the SET to allow for a possible tunnel event. Thus, the dark subspace readout effectively probes all remaining nuclear states in a single wait time, eliminating the need to wait separately for each of the 8 states, which is a key advantage of the adaptive protocol.

In case the initial guess was correct, tunneling events should be absent in the dark state subspace measurement. By setting a tunneling event threshold - here we choose ${\rm th} = {\rm}\lceil N_{\rm AR}/2\rceil$, where $N_{\rm AR}$ is the number of dark subspace readout shots - we decide whether we accept the initial guess (if $N_{\rm blips}< {\rm th}$) or reject it. In case $N_{\rm blips}\ge{\rm th}$, we discard the initial guess and repeat the protocol. 

This strategy allows us to identify the nuclear state with, in principle, just a single electron tunneling event per nuclear spin readout, which significantly reduces measurement-induced spin flips compared to repeated QND readout. The results displayed in Fig.~\ref{fig:3}c-d confirm the readout improvement with the adaptive protocol. When we apply the adaptive protocol, we observe a clear reduction in both the $\Delta m=1$ (green) and $\Delta m=2$ (purple) transition probabilities, consistent with a reduction in the number of tunneling events that occurred throughout the protocol. 

Figure~\ref{fig:3}e shows a comparison between repeated readout (RR) and adaptive readout (AR), both with an optimized number of readout shots $N_{\rm RR}$ and $N_{\rm AR}$. We show that the AR fidelities are on par with the measured single-shot ionization shock probabilities (from Fig.~\ref{fig:2}c); the optimal datapoint at $N_{\rm AR}=2$ coincides with the y-intercept of the black dashed line, confirming that the AR protocol indeed requires only one tunneling event. Furthermore, the measured readout fidelities for the adaptive protocol in Fig.~\ref{fig:3}e-f are consistent with the simulated ones described in Section~\ref{sec:sim_imperfect_ancilla}. The deviation observed for RR can likely be attributed to day-to-day variations in the quadrupole tensor or the ancilla readout fidelity, which leads to differences in the transition matrix compared to Fig.~\ref{fig:2} and variations in the number of tunneling events due to dark counts.

In addition to achieving better readout fidelities, the AR protocol provides a threefold speedup compared to the repeated QND readout. In the AR protocol, detecting the first blip takes on average 4 shots, followed by dark subspace readout with $N_{\rm AR}=2$. Since the initial guess is rejected approximately 13\% of the time, the total average number of QND cycles of $\langle N_{\rm QND}\rangle_{\rm AR} = 1/(1-0.13)\times6\approx6.9$. 

In contrast, the RR protocol requires $N_{\rm RR}=3$ (Fig.~\ref{fig:3}f, orange dots) repeated over 8 states, giving $\langle N_{\rm QND}\rangle_{\rm RR}=3\times8=24$ QND cycles. Accounting for the longer dark subspace pulse duration in the AR sequence, the AR protocol achieves an effective speedup of about a factor of three.

\subsection{Simulations with Imperfect Ancilla Readout}
\label{sec:sim_imperfect_ancilla}

In our experiments, the ancilla readout fidelity is already high with a true-positive probability of $\SI{0.968\pm0.003}{}$ and a false-positive probability of $\SI{0.019\pm0.001}{}$ - extracted from the raw traces of Fig.~\ref{fig:2}. To verify the robustness of the AR protocol under less ideal ancilla readout conditions, we perform a Monte-Carlo simulation of both the RR and AR protocols using the experimentally obtained transition matrix (Fig.~\ref{fig:2}) as a function of the ancilla readout fidelity.

\begin{figure}[h]
    \centering
    \includegraphics[width=\linewidth]{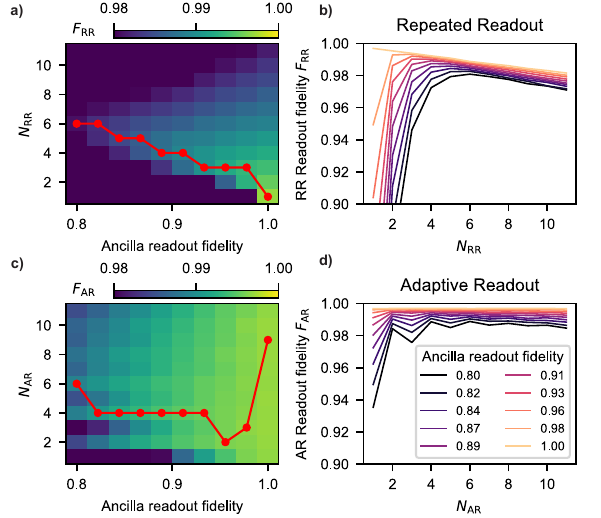}
    \caption{\textbf{Simulated comparison of repeated and adaptive QND readout protocols.}
    \textbf{a)} Optimal number of repeated readout shots (red line) increases as ancilla fidelity decreases.
    \textbf{b)} Maximum fidelity in repeated readout $F_{\rm RR}$ decreases with shot number $N_{\rm RR}$ due to non-ideal QND conditions.
    \textbf{c)} Adaptive readout achieves higher fidelities overall compared to repeated readout. When the ancilla readout fidelity approaches 1, the AR readout fidelity $F_{\rm AR}$ becomes insensitive to the number of shots $N_{\rm AR}$ due to the lack of tunneling events. The simulated optimal number of shots in that regime (red line) is subject to numerical fluctuations.
    \textbf{d)} Readout fidelity in adaptive readout remains robust against increasing dark space shots $N_{\rm AR}$, confirming suppression of measurement-induced errors. We ascribe oscillations in even-odd number of shots to the definition of the threshold ${\rm th =} \lceil N_{\rm AR}/2\rceil$, which slightly favors even $N_{\rm AR}$.}
    \label{fig:sim_readout_comparison}
\end{figure}

For the RR protocol (Fig.~\ref{fig:sim_readout_comparison}a) we find that a reduction in the ancilla readout fidelity requires an increase in the optimal number of shots (red line).
However, as the number of shots increases, so does the probability of measurement-induced transitions due to increased exposure to tunneling events. In Fig.~\ref{fig:sim_readout_comparison}b we find that, after reaching a maximum, the maximum readout fidelity gradually decreases as a function of the number of shots, illustrating the tradeoff between increased knowledge due to more shots and increased probability of a state transition due to non-ideal QND conditions.

In contrast, in the AR protocol (Fig.~\ref{fig:sim_readout_comparison}c), the fidelity remains relatively flat as a function of the number of dark subspace shots, confirming that the AR protocol indeed avoids measurement-induced transitions. Furthermore, we see overall higher readout fidelities compared to the RR case, where the fidelities are limited by the expected number of tunneling events, which slightly increases with imperfect ancilla readout fidelity due to dark counts. 

\subsection{Applicability to Other Systems: $^{73}$Ge with Pauli Spin Blockade readout}

\begin{figure}[h]
    \centering
    \includegraphics[width=\columnwidth]{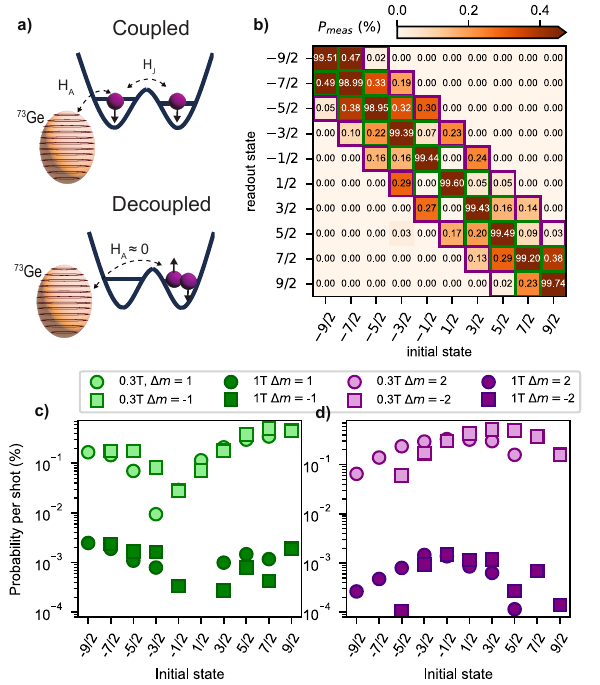}
    \caption{\textbf{Measurement-induced spin flips in a $\mathbf{\Ge}$ donor coupled to a double quantum dot.} \textbf{a)} Schematic of the system. \textbf{b)} Measurement-induced nuclear spin transition matrix for one ESR spectroscopy frequency sweep of 201 points. The transition matrix shows primarily $\Delta m =1$ (green) and $\Delta m = 2$ (purple) jumps, similar to backaction dynamics seen in the $\Sb$ system. \textbf{c--d)} Estimated transition probabilities per QND cycle. Increasing the magnetic field from $B_0 = \SI{0.3}{\tesla}$ (light) to $B_0 = \SI{1}{\tesla}$ (dark) reduces the transition probabilities.}
    \label{fig:4}
\end{figure}

To demonstrate that non-ideal QND conditions are a broader challenge in quantum devices, we consider an alternative system with a different readout mechanism. The second system under study is an isoelectronic $^{73}$Ge nuclear spin qudit in silicon, hyperfine-coupled to an electron in a quantum dot. Unlike the donor case, here the electron spin is measured using Pauli Spin Blockade (PSB) through a second, adjacent quantum dot \cite{Steinacker2025}. Although this system uses a different readout method than the $\Sb$ donor qudit, we observe similar nuclear spin flip signatures. This highlights that measurement-induced errors are a general feature that needs to be addressed across different platforms that employ non-ideal QND measurements.

We probe measurement-induced dynamics on the $\Ge$ system by performing ESR spectroscopy. 
We operate the device in a double quantum dot with a (9,1) charge configuration isolated from the electron reservoir, initialized in an odd parity electron spin state $\ket{\downarrow\uparrow}$.

After initialization, we apply an ESR pulse to flip the electron spin closest to the nucleus, while sweeping the ESR frequency across the expected resonance near the Larmor frequency $f_L=\gamma_eB_0$. We then read out the electron spin with PSB parity readout~\cite{Seedhouse2021}. The nuclear spin state $\ket{m}$ is inferred from the hyperfine-induced frequency shift $\Delta f = mA\approx m \times\SI{350}{\kilo \hertz}$~\cite{Steinacker2025}.
% with a $t_\mathrm{init} = \SI{10}{\micro \second}$ ramp across the (8,2) to (9,1) inter-dot charge transition. 

Fig.~\ref{fig:4}b shows the statistics of measurement-induced spin state transitions for all 10 eigenstates of the $\Ge$ nucleus. Similar to the $\Sb$ system, we observe both $\Delta m=\pm1$ and $\Delta m=\pm2$ measurement-induced nuclear spin flips. 
Each ESR frequency sweep produces one nuclear spin-state assignment, but it is important to note that the ESR measurement involves multiple tunneling events. Specifically, for every sweep we record 201 frequency points, and at each point the hyperfine-coupled electron is reinitialized through a tunneling-out and tunneling-in process, independent of the readout outcome. As a consequence, the system undergoes approximately $N_{\rm tunnel}=201$ electron tunneling cycles per sweep, each governed by a bare Markov transition matrix $T_{\rm Ge}$. The transition matrix shown in Fig.~\ref{fig:4}b therefore represents the compounded transition probabilites of 201 tunnel events, i.e. $T_{\rm Ge}$ raised to the power ${N_{\rm tunnel}=201}$. To extract the transition probabilities per QND cycle, we calculate the bare Markov matrix $T_{\rm Ge}$ (see Supplemental Materials Section~II~B for details), and display the spin flip probability per shot in Figure~\ref{fig:4}c~-~d.

We demonstrate a clear reduction of the transition probabilities for jumps of both $\Delta m=\pm1$ (green) and $\Delta m=\pm2$ (purple) from $B_0=\SI{0.3}{\tesla}$ (light colors) to $B_0=\SI{1}{\tesla}$ (dark colors). The reduction of spin flips with an increased magnetic field can be explained by the eigenstate overlap of the coupled and decoupled qudit. If the magnetic field is increased, the relative energies of the hyperfine and quadrupole interactions are reduced with respect to the Zeeman energy. As a consequence, eigenstates are more polarized along the magnetic field direction at high magnetic fields, and the overlap of the coupled and decoupled eigenstates is increased.

\section{Discussion and Conclusion}

By adaptively minimizing tunneling events, the demonstrated readout protocol outperforms conventional repeated QND readout in both fidelity and measurement time. These improvements make it a timely resource to improve the performance and fault-tolerance \cite{gottesman2016quantum} of QEC methods where a logical qubit is encoded in the large Hilbert space of a nuclear qudit \cite{Gross2021,Gross2024,Gupta2024,Yu2025}.

Recent work by Yu et al. \cite{Yu2025} demonstrated that the $\Sb$ nucleus can be used to encode a logical qubit in the spin-cat states of the form $\ket{\psi_{\rm cat}}_\Theta = 1/\sqrt{2}(\ket{-7/2}_\Theta +\ket{7/2}_\Theta)$, i.e. two maximally separated basis states defined along an axis $\Theta$ in the generalized rotating frame. Fault tolerant error correction schemes require all aspects of operating the code -- gates, initialization, and syndrome extraction -- to exceed the fault tolerant threshold \cite{gottesman2016quantum}. 

The adaptive protocol in this work is compatible with syndrome extraction in the spin-cat code, which relies on reading out the symmetric subspaces $\{\ket{-5/2}, \ket{5/2}\}$, 
$\{\ket{-3/2}, \ket{3/2}\}$, 
$\{\ket{-1/2}, \ket{1/2}\}$ \cite{Gross2024}. Here we demonstrate that we can increase the average QND readout fidelity to \SI{99.61\pm 0.04}{\percent}, a level consistent with estimates of error thresholds needed for fault tolerance \cite{Stephens2014, Ustun2024}. The improved readout fidelity is promising for reaching beyond the break-even for spin-cat codes. 

The added time efficiency becomes important when considering systems that have a long measurement time compared to their dephasing time $T_2$. Ideally, multiple rounds of error correction must be completed before the system decoheres. In the $\Sb$ system, the readout time (on the order of milliseconds) is still comparable to the free induction decay time $T_2^*$ \cite{Yu2025}. This limitation can only be partially mitigated through refocusing or qudit dynamical decoupling techniques \cite{Tripathi2025QuditProcessor}, which extend coherence but do not address the slow readout process. The second example, $\Ge$, offers a promising alternative by interfacing nuclear spins with gate-defined quantum dots, enabling PSB readout, which drastically reduces the measurement time~\cite{Geng2025} (\SI{100}{}--\SI{200}{\micro \second})~\cite{Steinacker2025} and allows for low-field as well as high-temperature operation~\cite{Huang2024High-fidelityK}. 
\FloatBarrier

\section*{Data and code availability}

All data and code to support the claims in the text will be made available after publication in an open online repository.

\section*{Declaration of interests}

A.M. is an inventor on a patent related to this work, describing the use of high-spin donor nuclei as quantum information processing elements (application no. AU2019227083A1, US16/975,669, WO2019165494A1). A.S.D. is the CEO and a director of Diraq Pty Ltd.. F.E.H., C.H.Y., A.S.D., A.L. declare equity interest in Diraq Pty Ltd.

\section*{Acknowledgments}
The research was funded by Australian Research Council Discovery Projects (grant no. DP210103769 and DP250101806) and the US Army Research Office (contract no. W911NF-23-1-0113). A.M. acknowledges support of an Australian Research Council Laureate Fellowship (project no. FL240100181).
We acknowledge the facilities, and the scientific and technical assistance provided by the UNSW node of the Australian National Fabrication Facility (ANFF), and the Heavy Ion Accelerators (HIA) nodes at the University of Melbourne and the Australian National University. ANFF and HIA are supported by the Australian Government through the National Collaborative Research Infrastructure Strategy (NCRIS) program. Ion beam facilities employed by D.N.J. and A.M.J. were co-funded by the Australian Research Council Centre of Excellence for Quantum Computation and Communication Technology (Grant No. CE170100012).
X.Y., B.W., M.R.v.B., A.V., P.S. acknowledge support from the Sydney Quantum Academy. P.S. acknowledges support from the Baxter Charitable Foundation. The views and conclusions contained in this document are those of the authors and should not be interpreted as representing the official policies, either expressed or implied, of the Army Research Office or the U.S. Government. The U.S. Government is authorized to reproduce and distribute reprints for Government purposes notwithstanding any copyright notation herein.

\section*{Author contributions}   
A.V. performed the experiment, data analysis and readout simulations. R.Y.S. wrote the measurement code and helped with simulations. L.A.O\textquotesingle N. performed the quadrupole Hamiltonian simulations. P.S. and G.G. performed the $\Ge$ experiment under supervision of A.L. and A.S.D..
M.N., M.R.v.B, X.Y., B.W. gave experimental input and helped characterizing the device.
D.H. and F.E.H. fabricated the device on materials supplied by K.M.I.. 
A.M.J., D.H. and D.N.J. performed the ion implantation.
C.H.Y. performed the $\Ge$ data analysis.
A.M. supervised the project.
A.V. and A.M. wrote the manuscript with input from all other authors.
% A.V. and M.N. conducted the experiment with input from B.W., X.Y., D.H., M.v.B., D.S., A.K. and A.M.. D.H., S.Y. and B.W. performed the device characterization. L.H.Z. and N.A. gave theoretical inputs. A.V., M.N., B.W., X.Y., wrote the measurement and analysis code. D.H. and F.E.H. fabricated the device,
% with A.S.D.’s supervision, on materials supplied by K.M.I.. A.M.J, D.H. and D.N.J. designed and performed the ion implantation. R.J.M. and R. B.-K. developed the qudit tomography protocol. A.V., M.N., L.H.Z. and A.M. wrote the manuscript with input from the other authors. A.M. supervised the experiments, V.S. supervised the theory development.

%\section{Competing interests}
%A.M. is an inventor on a patent related to this work, describing the use of high-spin donor nuclei as quantum information processing elements (application no. AU2019227083A1, US16/975,669, WO2019165494A1). A.S.D. is the CEO and a director of Diraq Pty Ltd.. F.E.H., A.L. and A.S.D. declare equity interest in Diraq Pty Ltd. All other authors declare no competing interests.

% \section{Additional information}

% TC:endignore
\bibliographystyle{unsrtnat}
\bibliography{bib}

\end{document}

\title{Supplemental Material: Maximizing the nondemolition nature of a quantum measurement via an adaptive readout protocol}

% \title{Minimizing the impact of non-ideal QND measurements with an adaptive readout protocol}

% \title{Minimizing measurement-induced errors with an adaptive readout protocol}

\author{Arjen Vaartjes}
\affiliation{School of Electrical Engineering and Telecommunications, UNSW Sydney, Sydney, NSW 2052, Australia}
\author{Rocky Yue Su}
\affiliation{School of Electrical Engineering and Telecommunications, UNSW Sydney, Sydney, NSW 2052, Australia}
\author{Laura A. O'Neill}
\affiliation{School of Electrical Engineering and Telecommunications, UNSW Sydney, Sydney, NSW 2052, Australia}
\author{Paul Steinacker}
\affiliation{School of Electrical Engineering and Telecommunications, UNSW Sydney, Sydney, NSW 2052, Australia}
\affiliation{Diraq Pty. Ltd., Sydney, NSW, Australia}
\author{Gauri Goenka}
\affiliation{School of Electrical Engineering and Telecommunications, UNSW Sydney, Sydney, NSW 2052, Australia}
\author{Mark R. van Blankenstein}
\affiliation{School of Electrical Engineering and Telecommunications, UNSW Sydney, Sydney, NSW 2052, Australia}
\author{Xi Yu}
\affiliation{School of Electrical Engineering and Telecommunications, UNSW Sydney, Sydney, NSW 2052, Australia}
\author{Benjamin Wilhelm}
\affiliation{School of Electrical Engineering and Telecommunications, UNSW Sydney, Sydney, NSW 2052, Australia}
\author{Alexander M. Jakob}
\affiliation{School of Physics, University of Melbourne, Melbourne, VIC 3010, Australia}
\author{Fay E. Hudson}
\affiliation{School of Electrical Engineering and Telecommunications, UNSW Sydney, Sydney, NSW 2052, Australia}
\affiliation{Diraq Pty. Ltd., Sydney, NSW, Australia}
\author{Kohei M. Itoh}
\affiliation{School of Fundamental Science and Technology, Keio University, Kohoku-ku, Yokohama, Japan}
\author{Chih Hwan Yang}
\affiliation{School of Electrical Engineering and Telecommunications, UNSW Sydney, Sydney, NSW 2052, Australia}
\affiliation{Diraq Pty. Ltd., Sydney, NSW, Australia}
\author{Andrew S. Dzurak}
\affiliation{School of Electrical Engineering and Telecommunications, UNSW Sydney, Sydney, NSW 2052, Australia}
\affiliation{Diraq Pty. Ltd., Sydney, NSW, Australia}
\author{David N. Jamieson}
\affiliation{School of Physics, University of Melbourne, Melbourne, VIC 3010, Australia}
\author{Martin Nurizzo}
\affiliation{School of Electrical Engineering and Telecommunications, UNSW Sydney, Sydney, NSW 2052, Australia}
\author{Danielle Holmes}
\affiliation{School of Electrical Engineering and Telecommunications, UNSW Sydney, Sydney, NSW 2052, Australia}
\author{Arne Laucht}
\affiliation{School of Electrical Engineering and Telecommunications, UNSW Sydney, Sydney, NSW 2052, Australia}
\affiliation{Diraq Pty. Ltd., Sydney, NSW, Australia}
\author{Andrea Morello}
\affiliation{School of Electrical Engineering and Telecommunications, UNSW Sydney, Sydney, NSW 2052, Australia}

\date{\today}

\maketitle
\tableofcontents
\newpage

\FloatBarrier

\FloatBarrier
\renewcommand{\thefigure}{S\arabic{figure}}
\setcounter{figure}{0}
\renewcommand{\theequation}{S\arabic{equation}}
\section{Measurement-induced transition model}
\FloatBarrier

\subsection{Determining $\mathbf{\Sb}$ Hamiltonian Parameters}
\FloatBarrier
\label{sec:hamiltonian_params}

Accurate modeling of the measurement-induced dynamics in the $\Sb$ system requires knowledge of all Hamiltonian parameters.
We determine the hyperfine interaction $A$ using standard ESR spectroscopy and find $A=\SI{97.5\pm2.2}{\mega\hertz}$. 

\FloatBarrier

\begin{figure}
    \centering
    \includegraphics[width=0.5\linewidth]{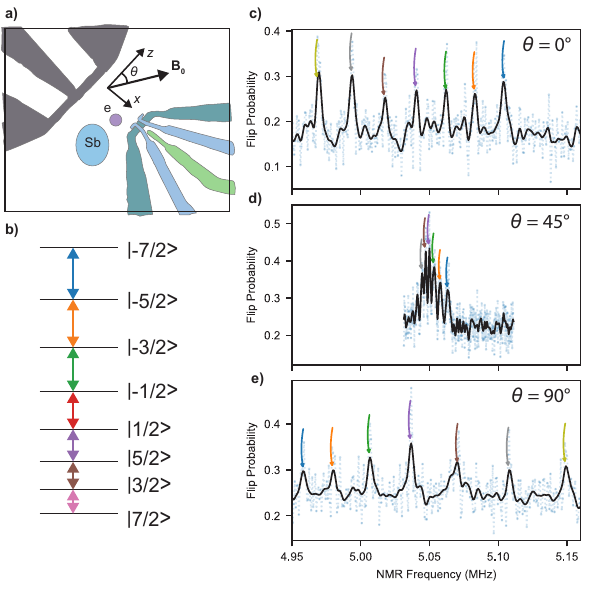}
    \caption{\textbf{Measurement to extract the quadrupole tensor.}
    \textbf{a)} Measurement setup. An $\Sb$ is implanted into purified $^{28}{\rm Si}$ and binds an electron, which is tunnel coupled to a Single Electron Transistor (SET). Four donor gates (not shown) control the potential landscape around the donor. The spin of the nucleus and the electron is controlled through a broadband antenna with Nuclear Magnetic Resonance (NMR) and Electron Spin Resonance (ESR), respectively. An external magnetic field $\mathbf{B_0}$ is applied in plane with the chip in the $xz$ coordinate system. 
    \textbf{b)} The $^{123}$Sb nucleus has 8 spin eigenstates which are separated the magnetic field. The anharmonicity is supplied by the nonlinear quadrupole interaction, which makes each NMR transition individually adressable.
    \textbf{c)} NMR spectra, where we identify 7 peaks corresponding to all $\Delta m=1$ spin transitions. The black line shows the gaussian filtered raw data (in blue). We track the peaks over various angles: $\theta=0^\circ$
    \textbf{d)} $\theta=45^\circ$. The separation between the peaks nearly vanishes.
    \textbf{e)} $\theta=90 ^\circ$. The order of the peaks has flipped.
    }
    \label{fig:rotated_nmr}
\end{figure}

We next characterize the quadrupole interaction in the ionized case. The quadrupolar part of the ionized $\Sb$ spin Hamiltonian depends on a $3 \times 3$ tensor $Q^+$, which is traceless and symmetric and therefore has five independent components \cite{Asaad2020, ONeill2021EngineeringSilicon}. Upon rotation of the magnetic field in the $zx$-plane (Fig.~\ref{fig:rotated_nmr}a), the Hamiltonian becomes angle-dependent:

\begin{equation}
    H_S = \gamma_n B_0\left(\cos{\theta}I_z + \sin{\theta}I_x \right) + \sum_{\alpha, \beta} Q^+_{\alpha\beta} I_\alpha I_\beta,
    \label{eq:angle_hamiltonian}
\end{equation}

\begin{figure}
    \centering
    \includegraphics[width=0.5\linewidth]{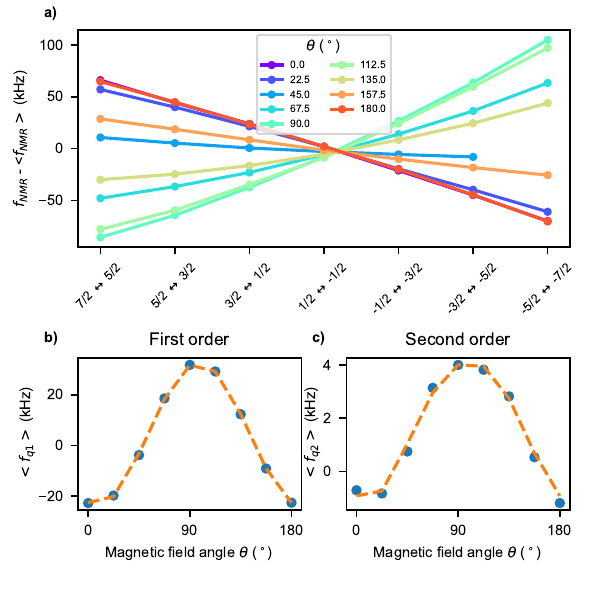}
    \caption{\textbf{Extracting the Quadrupole tensor. a)} Angular dependence of NMR frequencies $f_{\rm NMR}$, centered around the average $\langle f_{\rm NMR}\rangle$ for all 7 nuclear spin transitions in $\Sb$. \textbf{b)} Measured first-order quadrupole splitting $f_{q1}$, defined as the average difference between NMR frequencies. We observe a $180$ degrees periodicity characteristic to quadrupoles \cite{Franke2015}, and a sign change around $45$ degrees. \textbf{c)} Second order quadrupole splitting $f_{q2}$, defined as the difference of the average difference between NMR frequencies. Orange dashed lines show the model fit.}
    \label{fig:quadrupole}
\end{figure}

We perform Nuclear Magnetic Resonance (NMR) spectroscopy while rotating the magnetic field in-plane with the chip, and measure the angular dependence of both the first-order $f_{q1}=\langle \Delta f_{\rm NMR} \rangle$ and second-order $f_{q2}=\langle \Delta (\Delta f_{\rm NMR}) \rangle$ quadrupole splittings between the NMR frequencies $f_{\rm NMR}$ \cite{Franke2015}. Figure~\ref{fig:rotated_nmr} displays the rotated NMR method and indicates the coordinate system of the device. We fit the observed angular dependence of NMR frequencies to 5 independent parameters of the $Q^+$ tensor using the model described in Eq.~\ref{eq:angle_hamiltonian} (see Fig.~\ref{fig:quadrupole}). The best fit is obtained with the following quadrupole tensor elements:

\begin{table}[h]
    \centering
    \begin{tabular}{|c|c|c|}
    \hline
        Component & Value (\SI{}{\kilo\hertz}) & $\sigma$ (\SI{}{\kilo\hertz}) \\
        \hline
        $Q^+_{xx}$ & $-10.57$ & 0.17\\
        $Q^+_{yy}$ & $3.06$ & 0.20\\
        $Q^+_{zz} = -(Q^+_{xx}+Q^+_{yy})$ & $7.50$ & 0.15\\
        $Q^+_{yz}$ & $5.16$ & 0.35\\
        $Q^+_{xz}$ & $2.60$ & 0.45\\
        $Q^+_{xy}$ & $-30.48$ & 0.15\\

        \hline
    \end{tabular}
    \caption{Fitted values and standard errors of the ionized quadrupole tensor $Q^+$. The expression for $Q^+_{zz}$ follows from the traceless property of the tensor. }
    \label{tab:quadrupole_parameters}
\end{table} 
\FloatBarrier
The full tensor is given by:
\begin{equation}
Q^+ =
\begin{pmatrix}
-10.57 & -30.48 & 2.60 \\
-30.48 & 3.06 & 5.16 \\
2.60 & 5.16 & 7.50
\end{pmatrix}~\SI{}{\kilo\hertz}
\label{eq:quadrupole_tensor}
\end{equation}
The obtained tensor matches well with the measured angular dependence of the quadrupole splittings. At $\theta = 0$, we observe:
\[
\langle\Delta f^+_{\rm NMR}\rangle = \SI{-22.5}{\kilo\hertz}, \quad \langle\Delta^2 f^+_{\rm NMR}\rangle = \SI{700}{\hertz}.
\]

In the neutral state, we observe:
\[
\langle\Delta f^0_{\rm NMR}\rangle = \SI{-134}{\kilo\hertz}, \quad \langle\Delta^2 f^0_{\rm NMR}\rangle = \SI{906}{\hertz},
\]
which are consistent with the same quadrupole tensor $Q^+$. We therefore assume $Q^0 =Q^+$ in our Hamiltonian model (Eq.5 of the main text).

\subsection{Transition matrices for $\mathbf{\Sb}$}
\label{app:transition_matrix}

We attribute measurement-induced nuclear spin transitions to wavefunction collapse during Step 1 (coupling) and Step 3 (decoupling) in the QND readout cycle.

First, along the lines of Ref.~\cite{Monir2024}, we write the total Hamiltonian in an extended tensor basis formed by the $I_z$ nuclear spin eigenstates and the three possible electron spin configurations: coupled with spin up $\ket{\downarrow}$, coupled with spin down $\ket{\downarrow}$ or decoupled $\ket{\emptyset}$: $\{\ket{m}\} \otimes\{\ket{\uparrow}, \ket{\downarrow}, \ket{\emptyset}\}$. To expand the coupled (neutral) and decoupled (ionized) Hamiltonian to this extended tensor basis, we define two projection operators $P_{\updownarrow}$ and $P_\emptyset$~\cite{Monir2024}: 

\begin{equation}
     P_{\rm \updownarrow}=\mathbb{I}_8\otimes
     \begin{pmatrix}
         1 & 0 \\
         0 & 1 \\
         0 & 0
     \end{pmatrix}
\end{equation}

\begin{equation}
     P_{\rm \emptyset}=\mathbb{I}_8\otimes
     \begin{pmatrix}
         0 \\
         0 \\
         1
     \end{pmatrix},
\end{equation}
where $\mathbb{I}_8$ represents the $D=8$ identity matrix. The Hamiltonian in the extended basis is obtained by:
\begin{equation}
\begin{aligned}
        \mathcal{H} = P_{\rm \updownarrow} (H_S+H_A+H_C)P^{\dagger}_{\rm \updownarrow} +
    P_{\rm \emptyset} H_S P^\dagger_{\rm \emptyset}
\end{aligned}
\end{equation}

Diagonalizing this $24\times24$ Hamiltonian leads to 3 manifolds of 8 eigenstates: those primarily aligning with $\ket{\downarrow}$, labeled $\ket{m^\downarrow}$, those aligning with $\ket{\uparrow}$, labeled $\ket{m^\uparrow}$ and the decoupled eigenstates $\ket{m^\emptyset}$.

For the coupling process (Step 1), a $\ket{\downarrow}$ electron is loaded onto the nucleus from the SET. The associated nuclear spin flip probabilities are described by a stochastic transition matrix with elements:

\begin{equation}
    T^{\rm couple}_{mn} = |\langle n^\emptyset|P_{\emptyset}P^\dagger_\downarrow|m^\downarrow\rangle|^2 + |\langle n^\emptyset|P_\emptyset P^\dagger_\uparrow|m^\downarrow\rangle|^2,
    \label{eq:coupling_matrix}
\end{equation}

Here, the operators $P^\dagger_\downarrow=(\mathbb{I}_8\otimes\ketbra{\downarrow}{\downarrow})^\dagger$ and $P^\dagger_{\uparrow}=(\mathbb{I}_8\otimes\ketbra{\uparrow}{\uparrow})^\dagger$ project the eigenstates of the extended Hilbert space to the $\ket{\uparrow}$ and $\ket{\downarrow}$ subspaces, respectively. These projectors reduce the $24-$dimensional eigenvectors to the $8-$dimensional nuclear subspace. Including both $P^\dagger_\downarrow$ and $P^\dagger_\uparrow$ ensures that both the $\ket{\downarrow}$ and the (much smaller, flip-flop induced) $\ket{\uparrow}$ components of $|m^\downarrow\rangle$ are taken into account. The operator $P_{\emptyset}=\mathbb{I}_8\otimes\ketbra{\emptyset}{\emptyset}$ working on $\bra{n^\emptyset}$ projects the decoupled eigenstate to the corresponding nuclear subspace, reducing the length-24 vector to 8 as well. The projections allow calculating the nuclear overlap between coupled and decoupled Hamiltonian eigenstates.

In Step 3, if the electron flips to $\ket{\uparrow}$ by the aESR pulse, it tunnels out and the nucleus becomes decoupled from the ancilla again. The associated nuclear spin flip probabilities are described by:

\begin{equation}
    T^{\rm decouple}_{ij} = |\langle n^\emptyset|P_{\emptyset}P^\dagger_\uparrow|m^\uparrow\rangle|^2 + |\langle n^\emptyset|P_\emptyset P^\dagger_\downarrow|m^\uparrow\rangle|^2,
    \label{eq:decoupling_matrix}
\end{equation}

Because the two transition matrices are stochastic Markov matrices, we can describe the overall readout and initialization process $T_{\rm in-out}$, consisting of a tunnel-in and a tunnel-out event by simply multiplying the coupling and decoupling matrices:
\begin{equation}
    T_{\mathrm{in-out}} = T^{\mathrm{decouple}}T^{\mathrm{couple}}.
    \label{eq:qnd_transition_matrix}
\end{equation}
As discussed in Section~\ref{sec:num_tunneling}, multiple tunneling events can occur within a single readout sequence. To reproduce the experimentally observed transition matrix in Fig.~2c of the main text, we should therefore consider repeated multiplications of $T_{\rm in-out}$.

\begin{figure}
    \centering
    \includegraphics[width=0.5\linewidth]{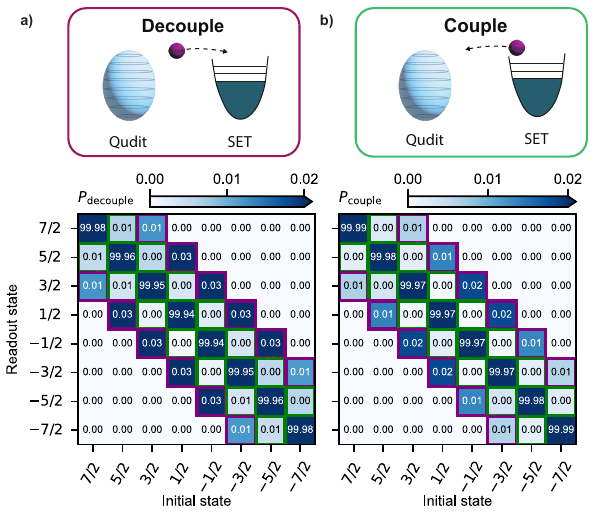}
    \caption{\textbf{Calculated transition matrices per QND cycle. a)} for decoupling ($T^{\rm decouple}$), we observe a slightly larger spin flip probability due to the decreased eigenstate overlap as discussed in Section~IIC of the main text. $\Delta m=\pm1$ and $\Delta m=\pm2$ spin transitions are indicated with green and purple squares.
    \textbf{b)} The transition matrix for coupling ($T^{\rm couple}$) the ancilla.}
    \label{fig:markov_matrices}
\end{figure}

Note that evaluating the transition probabilities via direct eigenstate projection corresponds to steady-state limit ($t \to \infty$) of the Lindblad equation describing the tunneling dynamics~\cite{Monir2024, Joecker2024}. This approach implicitly assumes that the readout window is much longer than the characteristic tunneling times. Given the high ancilla readout contrast observed experimentally, this assumption is well justified, indicating that tunneling events (blips) are almost never missed within a readout window. The calculated transition matrices per QND cycle are plotted in Fig.~\ref{fig:markov_matrices}.
This steady-state model leaves out more subtle time-dependent processes like flip-flop relaxation \cite{Savytskyy2023}, which were not characterized in this work.

\FloatBarrier
\section{Additional Data}
\FloatBarrier

\subsection{Number of tunnel events per QND cycle in $\mathbf{\Sb}$}
\label{sec:num_tunneling}
Readout and initialization of the electron spin is performed through the Maxwell's Demon protocol outlined in Ref.~\cite{Johnson2022}. This protocol boosts the confidence of initializing a $\ket{\downarrow}$ electron by applying a Bayesian update framework, essentially waiting a certain amount of time until the confidence of initializing a $\ket{\downarrow}$ electron exceeds a chosen threshold. If instead a spin $\ket{\uparrow}$ is loaded onto the nucleus, the electron will rapidly tunnel off, and the initialization protocol resets itself. 

\begin{figure}
    \centering
    \includegraphics[width=0.5\linewidth]{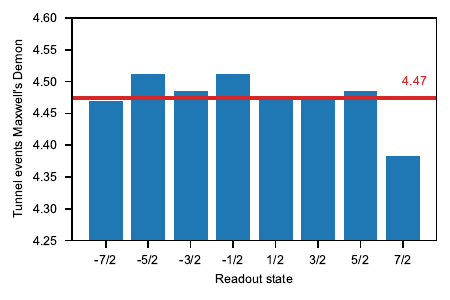}
    \caption{\textbf{Number of tunneling events per shot.} The Maxwell's demon readout and initialization protocol allows for multiple tunneling events per nuclear readout cycle \cite{Johnson2022}. We observe, on average, 4.47 tunneling events per QND readout cycle.}
    \label{fig:tunnel_maxwell_demon}
\end{figure}

Because the protocol continues until a $\ket{\downarrow}$ is initialized with high-confidence, it inherently allows multiple tunneling events within a single QND readout cycle. To quantify this effect, we performed a measurement in which each of the 8 nuclear spin states was prepared and read out with a single QND cycle (Fig.~2a). Using FPGA-based event tracking, we find that an electron tunnels on and off the SET an average of 4.47 times per QND cycle (see Fig.~\ref{fig:tunnel_maxwell_demon}).

We take these multiple tunneling events into account by raising the product of the bare Markov matrices to the power 4.47. 

\begin{equation}
    T_{\mathrm{QND}} = \left(T_{\mathrm{decouple}}T_{\mathrm{couple}}\right)^{4.47}
\end{equation}

The resulting transition matrix $T_{\rm QND}$ is shown in Fig.~2d of the main text.

\FloatBarrier

\subsection{Inferred bare Markov matrices for the $\Ge$ qudit}
\label{app:markov_ge}

In this section, we explain how to go from the cumulative transition matrix $T_{\rm Ge}^{N_{\rm tunnel}}$ (shown in Fig.~5b) to the bare Markov matrix $T_{\rm Ge}$, which describes the jump probabilities per readout shot shown in Fig.~5c--d.  

To find $T_{\rm Ge}$, we model the nuclear spin jumps as a Markov chain \cite{MarkovChains}. The bare Markov matrix can then be modeled as an exponential of a generator matrix $G$:
\begin{equation}
    T_{\rm Ge}^{N_{\rm tunnel}}=e^{N_{\rm tunnel} G},
    \label{eq:T_ge_exponential}
\end{equation}
where 
\begin{equation}
    G = \frac{1}{N_{\rm tunnel}}\log\left(T^{N_{\rm tunnel}}\right)
\end{equation}
is a continuous-time transition matrix. The single-shot Markov matrix can then be found with:
\begin{equation}
    T_{\rm Ge} = e^G.
\end{equation}

\begin{figure}
    \centering
    \includegraphics[width=0.5\linewidth]{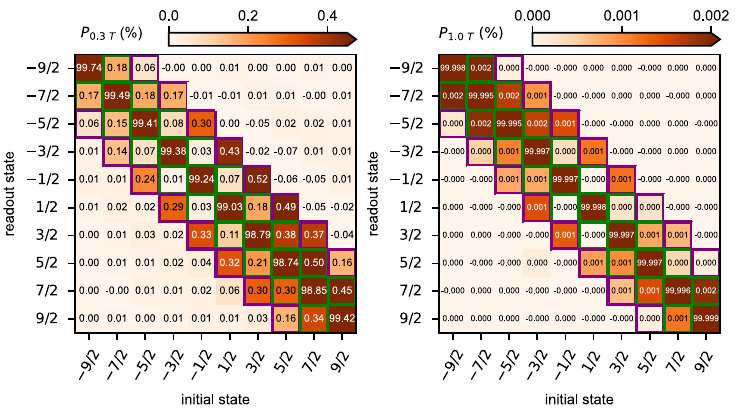}
    \caption{\textbf{Inferred Markov matrices per PSB readout shot of the $\Ge$ qudit. a)} For $B_0=\SI{0.3}{\tesla}$ and 
    \textbf{b)} $B_0=\SI{1.0}{\tesla}$. Note that the colorbar range is much smaller here.}
    \label{fig:markov_ge}
\end{figure}

Figure~\ref{fig:markov_ge} shows the calculated single-shot Markov matrices for $B_0=\SI{0.3}{\tesla}$ and $B_0=\SI{1.0}{\tesla}$. For $B_0=\SI{1.0}{\tesla}$, the $\Ge$ qudit is much more resilient against measurement-induced spin flips. The first and second off-diagonals of these matrices are plotted in Fig.~4c-d of the main text.

\FloatBarrier

\section{Experimental setups}\label{methods:Ge_experimental_device}

In this experiment we used two different qudit devices: a $\Sb$ donor and an isoelectronic $\Ge$, both implanted into isotopically enriched $\Sitwoeight$. The $\Sb$ device and measurement setup are nominally identical to those described in Ref.~\cite{Yu2025}, except for the ion implantation parameters. We used an implantation energy of \SI{10}{\kilo e \volt} and an ion fluence of \SI{4e11}{\centi\meter^{-2}} here. The $\Ge$ device and setup are described in detail in Ref.~\cite{Steinacker2025}, for which we used ion implantation parameters of \SI{12}{\kilo e \volt} and
a fluence of \SI{4e13}{\centi\meter^{-2}}.

\bibliographystyle{unsrtnat}
\bibliography{bib_sup}